\documentclass{camera}

\usepackage{graphicx}
\usepackage{epsfig}

\begin{document}

\title{Search for charged Higgs bosons in L3}
\author{P. Garcia-Abia}
\organization{CERN, EP Division, and University of Basel (Switzerland)}
\maketitle

\def\GeV{\ifmmode {\mathrm{\ Ge\kern -0.1em V}}\else
                   \textrm{Ge\kern -0.1em V}\fi}%
\def\pb{\mbox{pb$^{-1}$}}
\def\ra{\rightarrow}
\def\Hp{\ensuremath{\mathrm{H^+}}}
\def\Hm{\ensuremath{\mathrm{H^-}}}
\def\Hpm{\ensuremath{\mathrm{H^{\pm}}}}
\def\Wp{\ensuremath{\mathrm {W^+}}}
\def\Wm{\ensuremath{\mathrm {W^-}}}
\def\nbar{\ensuremath{\bar{\nu}}}
\def\ee{\ensuremath{\mathrm{e^+ e^-}}}
\def\antibar#1{\ensuremath{#1\bar{#1}}}%
\def\qqbar{\antibar{\mathrm{q}}}
\def\tntn{\mathrm{\tau^+\nu_\tau\tau^-\nbar_\tau}}
\def\cstn{\mathrm{c\bar{s} \tau^-\nbar_\tau}}
\def\cscs{\mathrm{c\bar{s} \bar{c}s}}
\def\qqen{qqe\nu_e}
\def\HHtntn{\mathrm{\Hp\Hm \ra \tntn}}
\def\HHcstn{\mathrm{\Hp\Hm \ra \cstn}}
\def\HHcscs{\mathrm{\Hp\Hm \ra \cscs}}
\def\l{\ifmath{\mathrm{\ell}}}
\def\EEHH{\ee\ra\Hp\Hm}
\def\Htn{\mathrm{H^\pm\ra\tau\nu}}
\def\BRTN{\mathrm{Br(H^\pm\ra\tau\nu)}}
\def\MHPM{\mathrm{m_\Hpm}}

\newcommand {\NP}     {Nucl.{} Phys.{} }
\newcommand {\PRL}    {Phys.{} Rev.{} Lett.{} }
\newcommand {\PL}     {Phys.{} Lett.{} }
\newcommand {\PR}     {Phys.{} Rev.{} }
\def\etal{{\it et~al.}}%
%
%

\begin{abstract}
  
  A search for pair-produced  charged Higgs bosons is performed with the
  L3 detector at LEP using data  collected  at  centre-of-mass  energies
  between 200 and 209~\GeV{},  corresponding to an integrated luminosity
  of  217.8~\pb.  We  observe an excess of events in the  $\HHcscs$  and
  $\cstn$  channels in the mass region around  68~\GeV{},  which is most
  significant  at low values of the $\Htn$  branching  ratio.  Including
  data taken at lower centre-of-mass  energies, the excess is compatible
  with a $4.4\sigma$  fluctuation in the background.  Interpreting  this
  excess as a statistical fluctuation in the background, lower limits on
  the charged Higgs mass are derived at the 95\% confidence level.  They
  vary from 67.1 to  84.9~\GeV{}  as a function of the $\Htn$  branching
  ratio.  These results are PRELIMINARY.

\end{abstract}


\section{Introduction}

In    the     Standard     Model~\cite{standard_model},     the    Higgs
mechanism~\cite{higgs_mech}  requires  one  doublet  of  complex  scalar
fields which leads to the  prediction of a single  neutral  scalar Higgs
boson.  Extensions to the minimal  Standard  Model contain more than one
Higgs   doublet~\cite{higgs_hunter}.  In  particular,  models  with  two
complex Higgs doublets  predict two charged Higgs bosons.

A search for the process  $\mathrm{\ee\ra\Hp\Hm}$  is  performed  in the
decay channels $\mathrm{\Hp\Hm\ra}$ $\tntn$, $\cstn$\footnote{The charge
conjugate  reaction is implied  throughout  this  letter.}  and $\cscs$,
assumed to be the only possible decays.  This allows the  interpretation
of the results to be independent of the $\Htn$ branching ratio.


\section{Data Analysis}

The search for pair-produced charged Higgs bosons is performed using the
data  collected  in  2000  with  the L3  detector~\cite{l3_det}  at LEP,
corresponding to an integrated luminosity of 217.8~\pb{} collected at an
average centre-of-mass energy of 205.9~\GeV{}.

The signature for the leptonic  decay  channel,  $\HHtntn$, is a pair of
tau leptons with large missing  energy and momentum,  giving rise to low
multiplicity  events with low visible energy and a flat  distribution in
acollinearity.  Almost all the background comes from W-pair  production.
The number of events  expected for a 70~\GeV{}  Higgs signal is 17.9 for
$\BRTN = 1$.

\begin{figure}[hp]
\centerline{\epsfig{figure=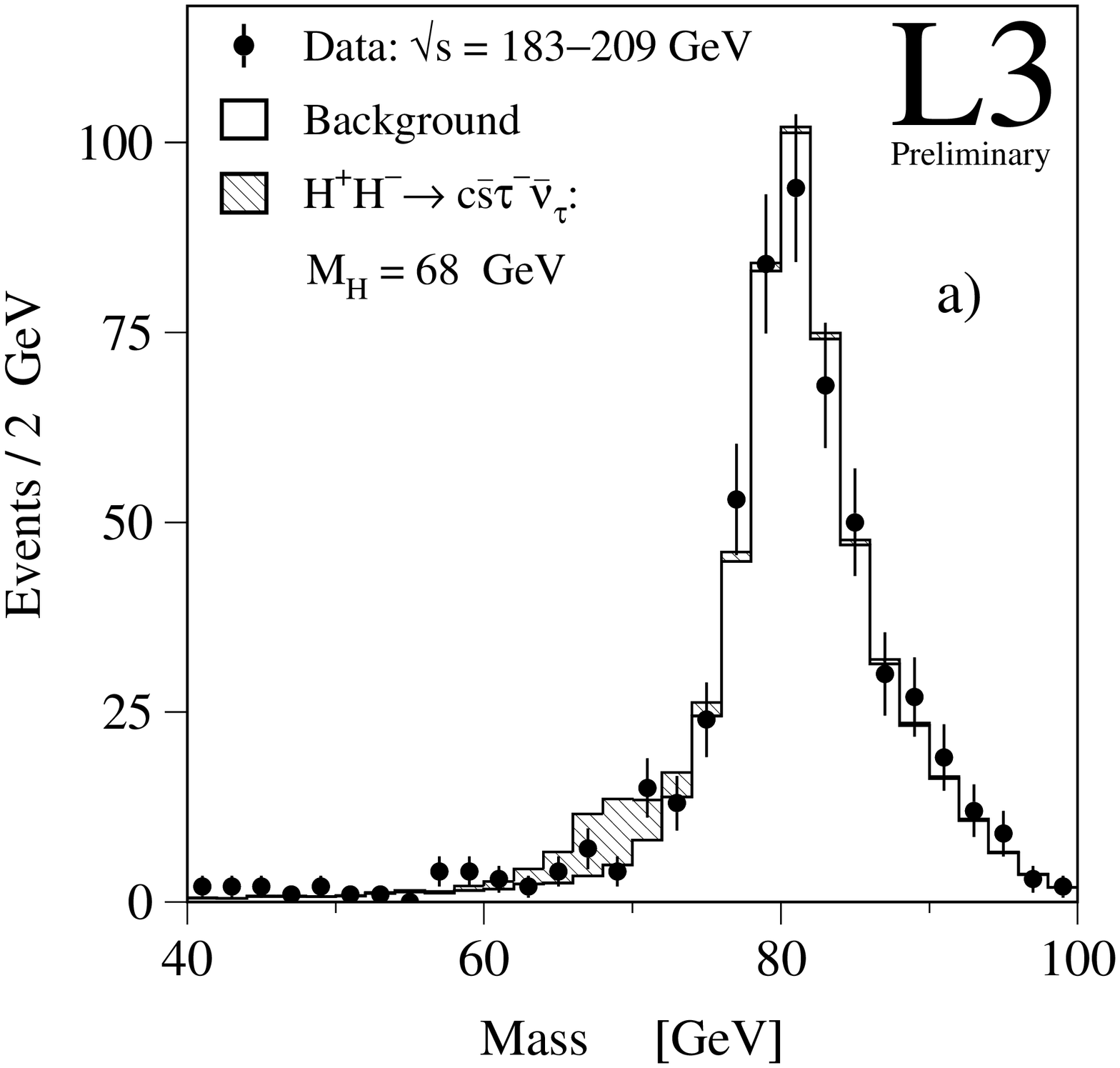,width=0.5\textwidth}
            \epsfig{figure=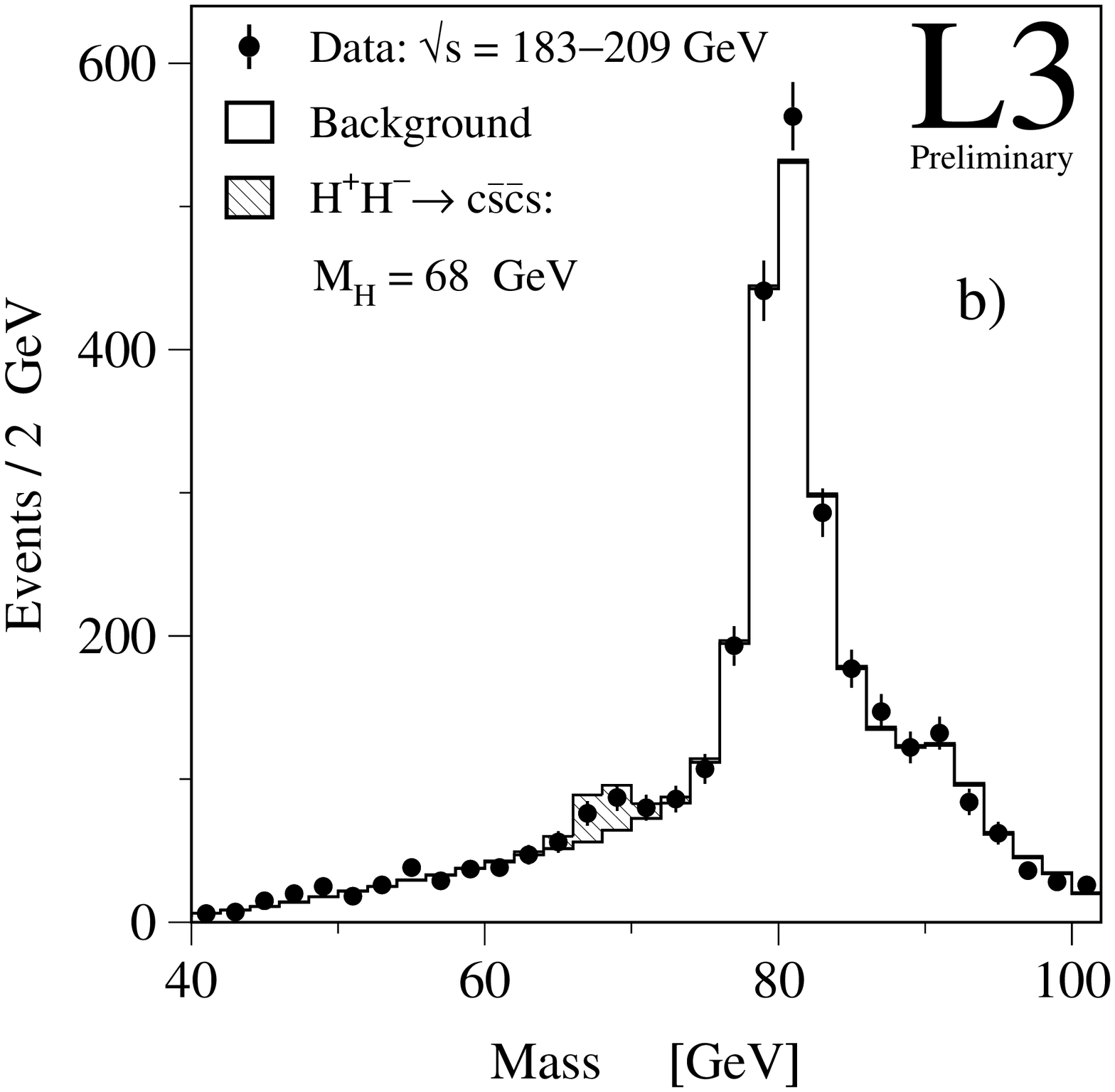,width=0.5\textwidth}}
\caption[]{\label{fig:massall}  Reconstructed  mass  spectra  in  the a)
  $\HHcstn$  and b)  $\cscs$  channels,  for  data and for the  expected
  background.  The hatched histogram indicates the expected distribution
  for a 68~\GeV{} Higgs with a)~$\BRTN = 0.5$ and b)~0.}
\end{figure}

The semileptonic  final state $\HHcstn$ is characterised by two hadronic
jets, a tau lepton and missing momentum.  The background is dominated by
the  process   $\Wp\Wm\ra\qqbar^\prime\tau\nu$.  The  number  of  events
expected  for a  70~\GeV{}  Higgs  signal  is 12.6  for  $\BRTN  = 0.5$.
Figure~\ref{fig:massall}a  displays the  distribution  of the average of
the  jet-jet  and  $\tau$-$\nu$  masses.  They  are  calculated  from  a
kinematic fit imposing energy and momentum  conservation  for an assumed
production of equal mass particles,  keeping the directions of the jets,
the tau and the missing momentum vector at their measured values.

Events  from the  $\HHcscs$  channel  have a high  multiplicity  and are
balanced in transverse  and  longitudinal  momenta.  A large fraction of
the  centre-of-mass  energy is deposited in the  detector,  typically as
four hadronic jets.  The main  contribution to the background comes from
W-pair  decays  into four  jets.  The  number of events  expected  for a
70~\GeV{}     Higgs    signal    is    35.1    for    $\BRTN    =    0$.
Figure~\ref{fig:massall}b  shows the  dijet  mass  distribution  after a
kinematic  fit  imposing  four-momentum  conservation  and  equal  dijet
masses.  An excess is observed around 68~\GeV{}.


\section{Results}

\begin{figure}[hp]
\centerline{\epsfig{figure=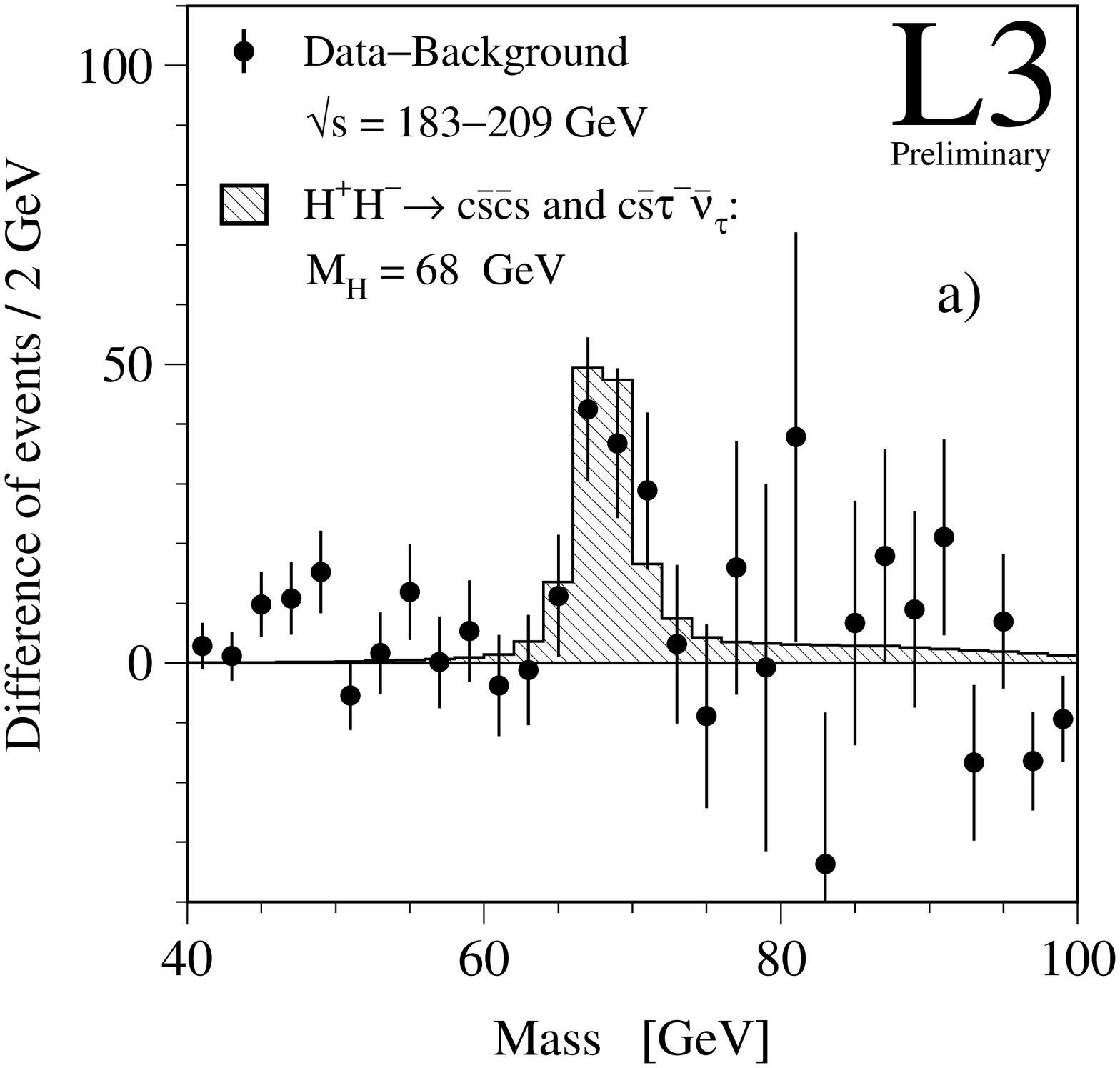,width=0.5\textwidth}
            \epsfig{figure=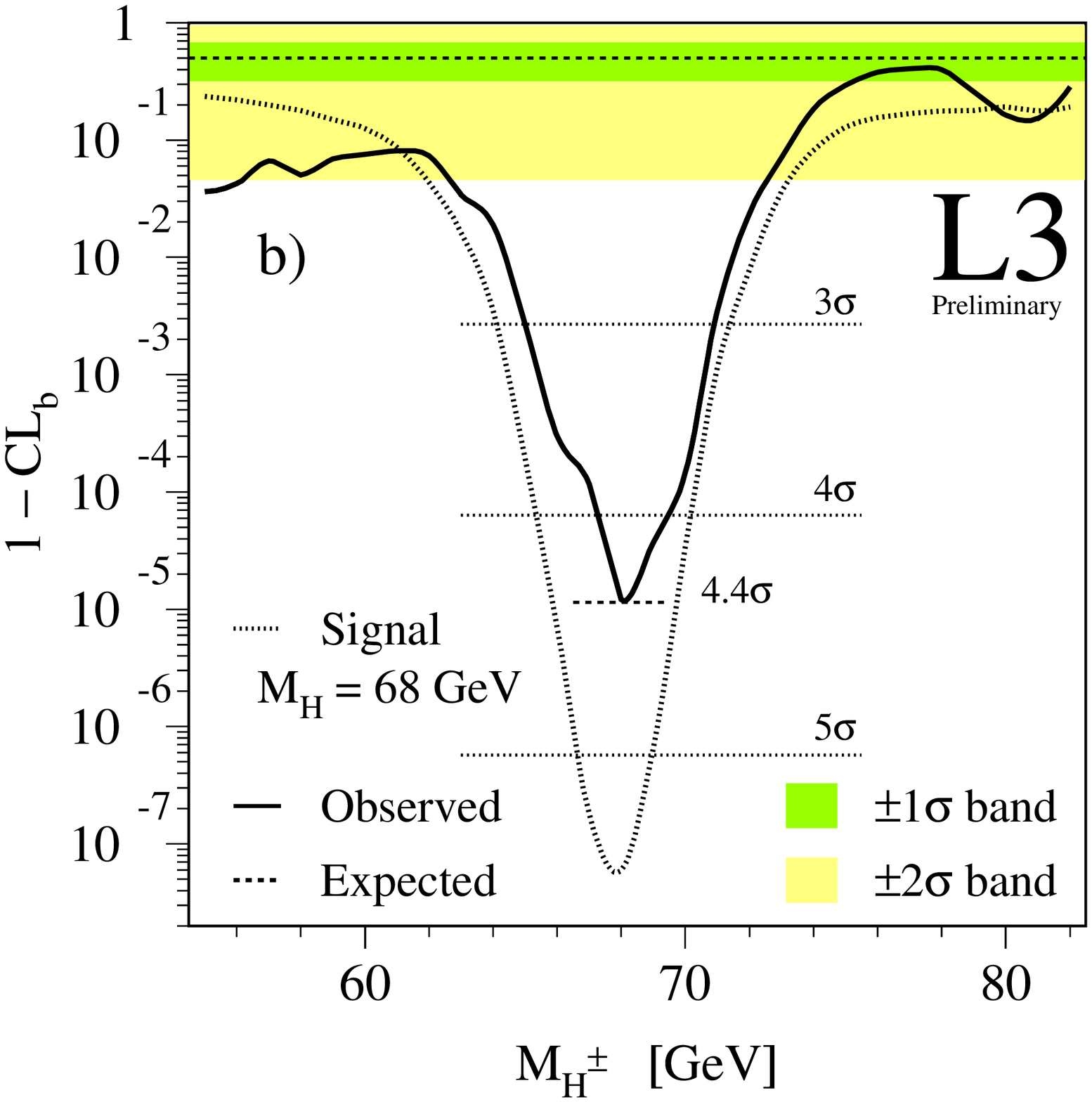,width=0.5\textwidth}}
\caption[]{\label{fig:mass}   a)  Combined   background-subtracted  mass
  distribution  for  the  $\HHcscs$  and  $\cstn$  decay  channels.  The
  expected  distribution  for a  68~\GeV{}  Higgs with  $\BRTN = 0.1$ is
  shown by the hatched histogram.
  b) the background  confidence level,  $1-\mathrm{CL_B}$, as a function
  of the Higgs mass with $\BRTN = 0.1$.  The solid line shows the values
  computed from the observed results and the dashed line the expectation
  for the  background  only  hypothesis.  The  dotted  line is the curve
  expected  for a 68~\GeV{} Higgs signal at this value of the  branching
  ratio.  The shaded  areas  represent  the  symmetric $1 \sigma$ and $2
  \sigma$ probability bands expected in the absence of a signal.}
\end{figure}

The number of selected  events in each decay channel is consistent  with
the number of events  expected from Standard Model  processes.  However,
there  is  an  excess  of  events  in  the  $\cscs$  and  $\cstn$   mass
distributions  around  68~\GeV{}.  Figure~\ref{fig:mass}a  displays  the
combined  background-subtracted  mass  distribution  for these two Higgs
decay  channels,  where the events are corrected  for the  efficiency of
their   respective   analyses.  The  figure  also  shows  the   expected
distribution  for a  68~\GeV{}  Higgs with $\BRTN = 0.1$.  This value of
the  $\BRTN$  is in the  range of  branching  fractions  for  which  the
observed  excess of events  is  closest  to the  expected  number  for a
68~\GeV{} mass Higgs.

A   technique    based   on   the    log-likelihood    ratio~\cite{lnq},
$-2\,\mathrm{ln}(Q)$,  is used to calculate a confidence level (CL) that
the observed  events are consistent with  background  expectations.  For
the $\cscs$ and $\cstn$ channels, the reconstructed  mass  distributions
(Figures~\ref{fig:massall}a  and~\ref{fig:massall}b)  are  used  in  the
calculation,  whereas for the $\tntn$ channel, the total number of data,
expected background and expected signal events are used.  The systematic
uncertainties  on the  background and signal  efficiencies,  5\% and 2\%
respectively, are included in the confidence level calculation.

The excess of events  around  $\MHPM = 68$~\GeV{} is  compatible  with a
$4.4\sigma$ fluctuation in the background.  The statistical significance
of the excess is almost  constant for values of $\BRTN$  between 0.1 and
0.2.  The data are  $1\sigma$  below what is expected for a Higgs signal
at this mass.  Again, this  difference is not strongly  dependent on the
value  of  the  branching  fraction.  The  background  confidence  level
($1-\mathrm{CL_B}$) is displayed in Figure~\ref{fig:mass}b for the data,
for the  expectation in the absence of a signal and for a 68~\GeV{} mass
Higgs signal.  Further  investigations  of the reported excess are being
performed.  Interpreting this excess as a statistical fluctuation in the
background,  lower limits on the charged Higgs mass as a function of the
$\BRTN$ are  derived~\cite{lnq,l3cl} at the 95\%~CL, using the data from
$\sqrt{s}$  between  199.6 and  209~\GeV{},  as well as those from lower
centre-of-mass  energies~\cite{chhiggs}.

Table~\ref{cl}  gives the observed and the median  expected lower limits
for several  values of the branching  ratio.  The region around $\MHPM =
68~\GeV$ at low values of the  $\BRTN$ can only be excluded  at 78\% CL,
due to the  aforementioned  excess of  events  in this  mass  region.  A
similar  but  less  significant  excess  was  observed  in our  previous
publication~\cite{chhiggs}.

\begin{table}
\begin{center}
\begin{tabular}{|c|cc|} \hline
\raisebox{-8pt}[0pt][0pt]{$\BRTN$}
      & \multicolumn{2}{|c|}{Lower limits at 95\% CL (\GeV{})} \\
      & \ observed \      & median expected                    \\ \hline
 0.0  &        77.2       &        77.1                        \\
 0.1  &        67.1       &        76.2                        \\
 0.5  &        70.4       &        75.6                        \\
 1.0  &        84.9       &        83.0                        \\ \hline
\end{tabular}
\caption{Observed  and  median  expected  lower  limits  at 95\%  CL for
    different   values  of  the  $\Htn$  branching  ratio.  The  minimum
    observed limit, independent of the branching  fraction, is at $\BRTN
    = 0.1$.}
\label{cl}
\end{center}
\end{table}

Our  sensitivity  to larger Higgs  masses, as  quantified  by the median
expected mass limits given in Table~\ref{cl}, is significantly  improved
as  compared   with  our  previous   results  at  lower   centre-of-mass
energies~\cite{chhiggs}.  Combining  all our data, we obtain a new lower
limit at 95\% CL of $\MHPM > 67.1 \GeV{},$  independent of the branching
ratio.  These results are PRELIMINARY.


\end{document}